\documentclass[12pt,epsf]{article}

\begin{document}
\title{Dimension-5 operators in a Gauge Mediated Supersymmetry Breaking Model}
\author{Da-Xin Zhang\footnote{dxzhang@mail.phy.pku.edu.cn}\\
\it\small Institute of Theoretical Physics,\\[-3mm]
\it\small School of Physics, Peking University, Beijing 100871,
China}
\date{}

\maketitle

\begin{abstract}
We study the novel features in a model with Gauge Mediated
Supersymmetry Breaking. If the messenger fields have positive
R-parity, there will be new sources of flavor violations. We show
that the dimension-5 operators will be quite important. When
dressing these operators by wino-loops, the constraints on them by
the present data are given.
\end{abstract}

\vspace{0.5cm} {\it PACS}: 12.15.Mm, 12.15Cc, 11.30Pb

\newpage
In models with Gauge Mediated Supersymmetry Breaking
(GMSB)\cite{gmsb,review}, there exist the so-called ``messenger
fields'' which transfer the information of supersymmetry (SUSY)
breaking from a hidden sector to the visible sector through gauge
interactions. The minimal model of this kind consists of
vector-like messenger fields ${\bf 5+{\bar 5}}$ under SU(5).
Variations of the minimal model in many directions have also been
studied carefully in the literature, resulting differences  both
in spectra and in phenomena of the Minimal Supersymmetric Standard
Model(MSSM) \cite{martin,kai}. Even in the the minimal model of
GMSB there can be direct interactions between the messenger fields
and the matter fields. It has been noticed that the messenger
fields can have either positive (Higgs-like) or negative
(matter-like) R-parities\cite{hanzhang,dub}. The former case has
very rich consequences in the low energy physics.

In the present work, we will focus on the minimal model of GMSB
with positive R-parity messengers. We will assume that R-parity is
conserved. We will study the special features of these
interactions between the messenger and the matter fields, and use
the low energy data to bound the allowed couplings.

It is interesting to note that in the GMSB models, the typical
masses for the messenger fields are $M_{mess}\sim 100$TeV. They
are heavier than the squarks/sleptons and gauginos/higgsinos
($\tilde m\sim 100$GeV $- 1$TeV)\cite{kai}, and much lighter than
the color-triplet Higgs multiplets of the unification theory.
Lessons on SUSY SU(5) grand unified theory\cite{mura} taught us
that the dimension-5 operators are usually the dominant sources
inducing nucleon decays, as they are less power suppressed by the
heavy colored Higgs(ino) mass than the dimension-6 operators
mediated by either gauge bosons or Higgs bosons. Furthermore,
since the dimension-5 operators will be dressed by -ino (gaugino
or higgsino) loops to be compared with the data,  results on
different flavor structures will be reached. Although these
effects are suppressed by the loop factors ({\it e.g.},
$1/(4\pi)^2$), they are compensated by the factor
$M_{mess}/{\tilde m}$ to compare with the scalar messenger
mediated 4-fermion interactions which are dimension-6. Note also
that the dimension-5 operators are not important in the study of
R-parity violating models due to the absence of this power
enhancement, while being mediated by colored Higgs which are very
heavy, they are negligibly small in the study of flavor physics in
unification theories.

In the case of positive R-parity messenger fields, the
superpotential between the messenger fields and the matter fields
is\cite{hanzhang}
\begin{eqnarray}
W&=& y_{ij} {\overline E_i} L_j L_4 \  +\ y'_{ij} {\overline D_i}
Q_j L_4 \ +\ y''_{ij} {\overline U_i} Q_j {\overline L_4}
\nonumber\\
&+& \lambda^l_{ij} {\overline E_i} {\overline U_j} D_4 \ +\
\lambda'_{ij} L_i Q_j {\overline D_4}\ +\ {1\over 2}\lambda_{ij}^q
Q_i Q_j D_4 \ +\ \lambda''_{ij} {\overline U_i} {\overline D_j}
{\overline D_4}\ .\label{w}
\end{eqnarray}
Here
\begin{eqnarray}
{\bf 5+{\bar 5}}=(D_4,{\overline L_4})+({\overline D_4},L_4)
\end{eqnarray}
are the messenger superfields. They have the same quantum numbers
as the ${\bf 5+{\bar 5}}$ Higgs of SU(5) theory. Bilinear terms
will not be discussed here. We will not discuss the soft breaking
messenger interaction either. The scalar messenger fields can
induce the 4-fermion interactions between the matter fields which
are dimension-6 and are proportional to $1/M_{mess}^2$.
Consequently the low energy data will bound many of the couplings
in (\ref{w})\cite{hanzhang}.

Now we focus on the dimension-5 operators  which are suppressed by
only $1/M_{mess}$. From (\ref{w}) these operators can be divided
into two classes. The first class of operators are relevant to the
nucleon decays and are described by (see Fig. 1)
\begin{eqnarray}
L_5^{(1)} & = & -\frac{1}{M_{D_4}} \left[ {1\over
2}\lambda_{ij}^q\lambda'_{lk}  (Q_i Q_j)( Q_k L_l) +
\lambda''_{ij}\lambda^l_{lk} (\bar U_i \bar D_j)( \bar U_k \bar
E_l)\right]. \label{pd}
\end{eqnarray}
This is more general than the SU(5) superpotential form
\cite{mura} due to the Higgs color-triplets. Because
$\lambda_{ij}^q$ is symmetric with respect to its two indices,
under the condition
\begin{eqnarray}
\lambda_{ij}^q\lambda'_{lk}=&\lambda_{ik}^q\lambda'_{lj}&=\lambda_{jk}^q\lambda'_{ik},\nonumber\\
\lambda''_{ij}\lambda^l_{lk}=&\lambda''_{ik}\lambda^l_{lj}&=\lambda''_{kj}\lambda^l_{li},
\label{hc}
\end{eqnarray}
the two terms in (\ref{pd}) can be combined into the same
superpotential form as that relevant to nucleon decays in SU(5).
 The second class of operators conserve both baryon number
and totally lepton numbers (see Fig. 2). Due to the fact that the
couplings in (\ref{w}) are independent, again  these interactions
are written as
\begin{eqnarray}
L_5^{(2)} & = & - \frac{\lambda'_{ij}\lambda^l_{kl}}{M_{D_4}}(L_i
Q_j)(\bar E_k\bar U_l)
  +\frac{y_{ik}y''_{jl}}{M_{L_4}}(L_i^a\bar E_k  )(Q_j^b \bar
  U_l)\nonumber\\&&
  - {1\over 2}\frac{\lambda^q_{ij} \lambda''_{kl}}{M_{D_4}}
   (Q_i Q_j)^\alpha (\bar U_k \bar D_l)^\alpha
 +\frac{y'_{ik}y''_{jl}}{M_{L_4}}\epsilon^{ab}
    (Q_i^a\bar U_k )(Q_j^b \bar D_l), \label{fc}
\end{eqnarray}
where $a,b=1,2$ are flavor indices, $\alpha=1,2,3$ is color index.
In (\ref{pd}) and (\ref{fc}) one fermion and one scalar are
selected in each bracket, {\it e.g.},
\begin{eqnarray}
(Q_i Q_j)^\alpha \equiv \epsilon^{\alpha\beta\gamma}(\tilde
u_i^\beta d_j^{\prime\gamma}-\tilde
  d_j^{\prime\gamma} u_i^\beta).\label{eg}
\end{eqnarray}
 Note that the analogue superpotential of (\ref{fc}) in SUSY
SU(5) does not violate separate lepton numbers and is negligibly
small due to the suppression factor $1/M_{H_C}$.

Supposing that the couplings in (\ref{w}) are  too small to affect
the spectrum of the MSSM with GMSB, they can still induce many
consequences in the low energy experiments. We will assume that
the leading flavor changing effects are due to charged
interactions and those in (\ref{pd},\ref{fc}). The dominant
effects from these dimension-5 operators are dressed by charginos.
We further simplify the discussion by taking the wino as the most
important component of the charginos. We get the relevant
Lagrangian
\begin{eqnarray}
\lefteqn{    {\cal L}^{(1)} =- \frac{1}{M_{D_4}}
\frac{\alpha_2}{2\pi} \lambda'_{ij}
\lambda^l_{kl}\epsilon^{\alpha\beta\gamma}
 }          \nonumber \\
& & \left[
    (d_i^{\prime\alpha} \nu_l) (d_j^{\prime\beta} u_k^\gamma)
                (f(u_i,\,e_l) + f(u_j,\,d'_k))
    + (u_i^\alpha e_l)(u_j^\beta d_k^{\prime\gamma})
                (f(u_k,\,d'_j) + f(d'_i,\,\nu_l)) \right. \nonumber \\
& & \left.
    + (d_i^{\prime\alpha} u_k^\beta)(d_j^{\prime\gamma} \nu_l)
                (f(u_i,\,d'_k) + f(u_j,\,e_l))
    + (u_i^\alpha d_k^{\prime\beta})(u_j^\gamma e_l)
                (f(u_k,\,d'_i) + f(d'_j,\,\nu_l))
                 \right]\nonumber \\&+ &{\rm h.c.}
\label{lpd}
\end{eqnarray}
for the first class of interactions which mediate nucleon decays.
All the fermions are left-handed, and we have taken $d^\prime$ as
the interaction eigenstate while $u$ as the mass eigenstate. Since
we are using the wino-dressing diagrams, the effects of the RRRR
operator in (\ref{pd}) are neglected. Note that the structure in
(\ref{lpd}) is different from that in the minimal SUSY SU(5), due
to the fact that the color-triplet Higgs interactions do not
violate flavors.

 After dressed by wino, the
second line in (\ref{fc}) results only nonleptonic 4-fermion
interactions which can hardly be as important as the W-mediated
interactions, while only the first line in (\ref{fc}) gives rise
to flavor changing neutral current (FCNC) interactions. We
concentrate on the FCNC interactions and get the Lagrangian
\begin{eqnarray}
{\cal L}^{(2)} &=&- \frac{\alpha_2}{2\pi}
\frac{y_{ik}y''_{jl}}{M_{L_4}}
f(\nu_i,\,d_{j'})
V_{jj'}V_{j''j'}^*\bar e_k \frac{1-\gamma_5}{2}u_l^c \bar
u_{j''}^c \frac{1-\gamma_5}{2}e_i + {\rm h.c.}\label{lfc}
\end{eqnarray}
which mediates rare and lepton number violating decays of the
up-type quarks (charm quark, especially) and of the $\tau$ lepton.
Note that the ($L\times L$) Lorentz structure  in (\ref{lfc}) is
different from the ($L\times R$) structure in the case of
dimension-6 operators. The function $f(f_1,f_2)$
\begin{equation}
         f(f_1,\,f_2) \equiv
        \frac{m_{\tilde{w}}}{m_{\tilde{f_1}}^2 - m_{\tilde{f_2}}^2}
        \left( \frac{m_{\tilde{f_1}}^2}{m_{\tilde{f_1}}^2 -m_{\tilde{w}}^2}
            \ln \frac{m_{\tilde{f_1}}^2}{m_{\tilde{w}}^2}
        - \frac{m_{\tilde{f_2}}^2}{m_{\tilde{f_2}}^2 - m_{\tilde{w}}^2}
            \ln \frac{m_{\tilde{f_2}}^2}{m_{\tilde{w}}^2} \right)
\label{f}
\end{equation}
is the standard triangle loop function.

We now use (\ref{lpd}) and (\ref{lfc}) to bound the couplings in
(\ref{w}).  We take the typically values
\begin{eqnarray}
M_{\tilde{w}}=500{\rm GeV}, ~M_{\tilde{f}}=800{\rm GeV},
~M_{D_4}=M_{L_4}=100{\rm TeV}\label{mass}
\end{eqnarray}
and $\alpha_2(m_Z)=0.0335$ in our numerical estimation. First we
consider nucleon decays. We use the $\tau(N\to \nu K) > 0.86
\times 10^{32}$years \cite{pdg} to get
\begin{eqnarray}
|\lambda^q_{12} \lambda'_{l1}|<3.7\times
10^{-21}\left(\frac{M_{D_4}}{100{\rm TeV}}\right) ~(l=1,2,3),
\label{bd1}
\end{eqnarray}
which is of the same order as the bounds on the other products of
couplings in (\ref{w}) ({\it e.g.}, $\lambda'\lambda''<10^{-21}$
in \cite{hanzhang}).

In the FCNC processes induced by the dimension-5 operators, the
appearance of  $V_{j'j}V_{j''j'}^*$ in (\ref{lfc}) indicates the
GIM cancellations among down-type squarks in the loops of Fig. 3.
Consequently, most of the results depend strongly on the details
of the model and the parameters which will not be discussed.
However, processes with $j''=j$ will not suffer the GIM mechanism
and we get
\begin{eqnarray}
y_{31(13)}y''_{11}<0.064\left(\frac{M_{L_4}}{100{\rm TeV}}\right)&&{\rm from} ~Br(\tau\to e \pi^0)<2.2\times 10^{-6}, \nonumber\\
y_{32(23)}y''_{11}<0.066\left(\frac{M_{L_4}}{100{\rm TeV}}\right)&&{\rm from}~Br(\tau\to \mu \pi^0)<8.2\times 10^{-6},\nonumber\\
y_{11}y''_{12(21)}<0.35\left(\frac{M_{L_4}}{100{\rm TeV}}\right)&&{\rm from}~Br(D^+\to \pi^+ e^+e^-)<8.8\times 10^{-6}, \nonumber\\
y_{12(21)}y''_{12(21)}<0.21\left(\frac{M_{L_4}}{100{\rm TeV}}\right)&&{\rm from}~Br(D^+\to \pi^+ e^\pm\mu^\mp)<3.4\times 10^{-5},\nonumber\\
y_{22}y''_{12(21)}<0.15\left(\frac{M_{L_4}}{100{\rm
TeV}}\right)&&{\rm from}~Br(D^+\to \pi^+\mu^+\mu^-)<5.2\times
10^{-5}. \label{bd2}
\end{eqnarray}
Here very simple estimations on hadronic matrix elements have been
used based on chiral perturbation theory\cite{wise}. It is
interesting that the bounds on $yy''$ are absent if the
dimension-6 operators are studied. The bounds in (\ref{bd2}) are
also comparable in size with the bounds on $yy,y''y''$ given in
the dimension-6 case\cite{hanzhang}.

We have studied the effects of novel dimension-5 operators in the
minimal GMSB model with  direct messenger-matter interactions. We
find that these effects are comparable with the dimension-6
operators. Although we have limited in discussing the minimal GMSB
model, our results  can be easily extended to other SUSY models
with extra chiral multiplets: if they are not too heavy, the
dimension-5 operators have important effects in the low energy
phenomena.

 This work was supported in part
by the National Natural Science Foundation of China (NSFC) under
the grant No. 90103014 and No. 10205001, and by the Ministry of
Education of China.

\newpage
\textwidth 14.5cm\oddsidemargin -0.35in\evensidemargin 0.3in
\begin{figure}
\begin{picture}(1000,300)
\put(20,280){\line(1,-1){40}} \put(40,260){\vector(1,-1){2}}
\put(160,240){\line(1,1){40}}\put(180,260){\vector(-1,-1){2}}
\put(60,240){\line(1,0){100}}\put(110,235.6){X}
\put(80,236.7){$<$}\put(140,236.7){$>$}
\put(20,200){\line(1,1){40}} \put(40,220){\vector(1,1){2}}
\put(160,240){\line(1,-1){40}}\put(180,220){\vector(-1,1){2}}
\put(300,280){\line(1,-1){40}} \put(320,260){\vector(1,-1){2}}
\put(440,240){\line(1,1){40}}\put(460,260){\vector(-1,-1){2}}
\put(340,240){\line(1,0){100}}\put(390,235.6){X}
\put(360,235.7){$<$}\put(420,235.7){$>$}
\put(300,200){\line(1,1){40}} \put(320,220){\vector(1,1){2}}
\put(440,240){\line(1,-1){40}}\put(460,220){\vector(-1,1){2}}
\put(280,280){$\bar U_k$}\put(280,190){$\bar E_l$}
\put(490,280){$\bar U_i$}\put(490,190){$\bar D_j$}
\put(0,280){$Q_i$}\put(0,190){$Q_j$}
\put(210,280){$Q_k$}\put(210,190){$L_l$}
\put(80,250){$D_4$}\put(140,250){$\bar D_4$}
 \put(360,250){$D_4$}\put(420,250){$\bar D_4$}
\put(60,225){$\lambda^q$}\put(155,225){$\lambda'$}\put(340,225){$\lambda^l$}\put(435,225){$\lambda''$}
\end{picture}
\vspace{-6.5cm} \caption{Dimension-5 operators which are relevant
to nucleon decays. }
\end{figure}
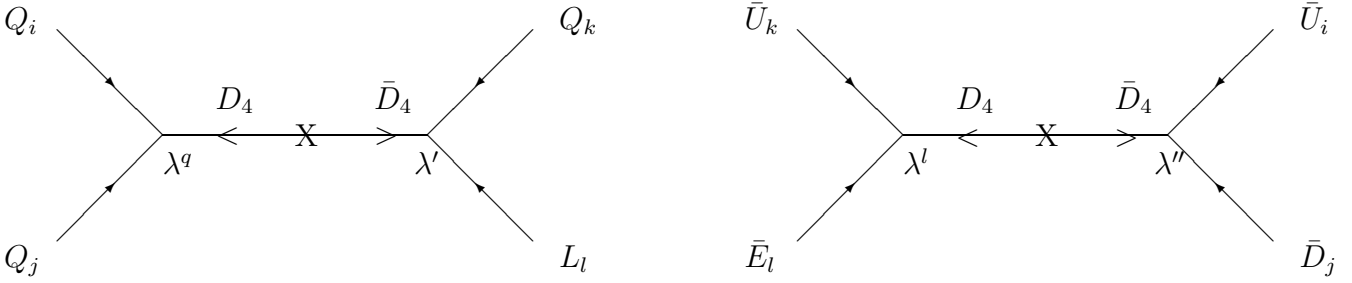

\vspace{-8.5cm}
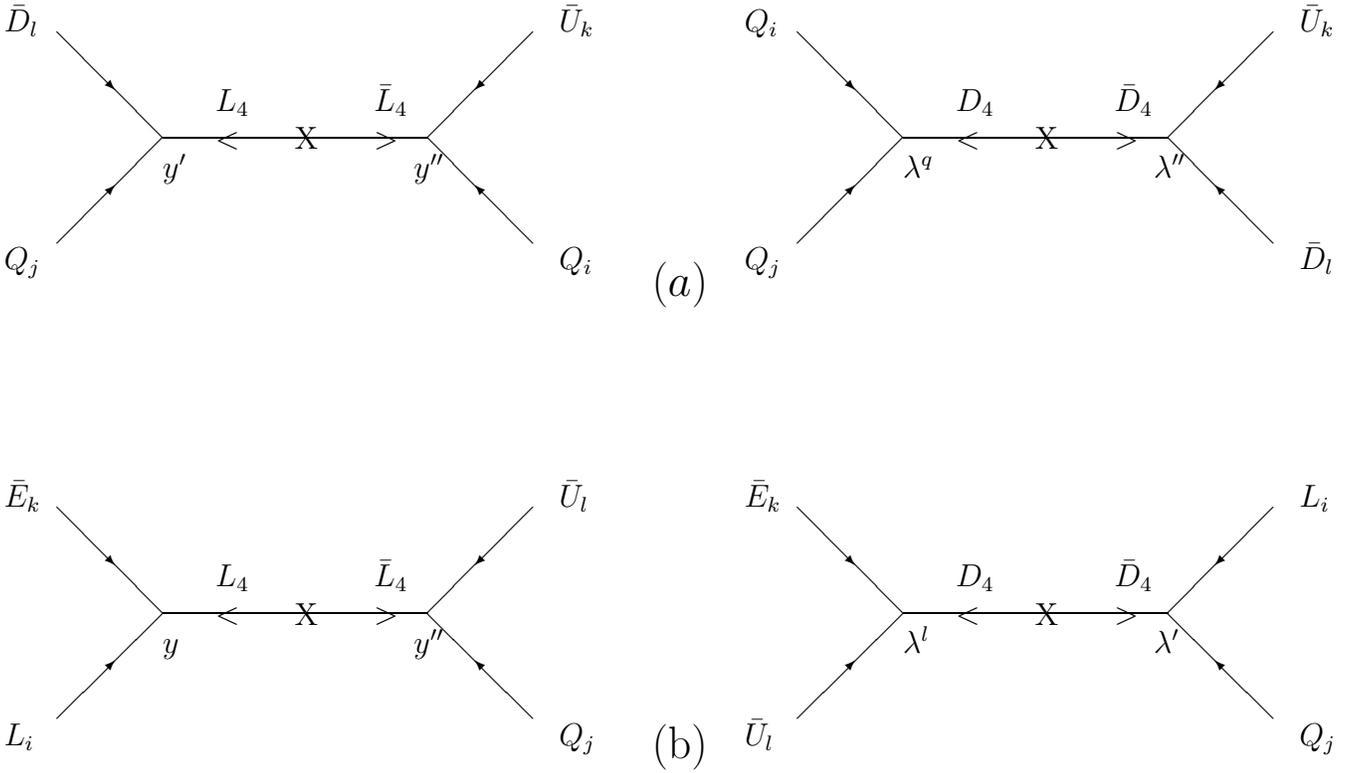
\begin{figure}
\begin{picture}(600,300)
\put(20,280){\line(1,-1){40}} \put(40,260){\vector(1,-1){2}}
\put(160,240){\line(1,1){40}}\put(180,260){\vector(-1,-1){2}}
\put(60,240){\line(1,0){100}}\put(110,235.6){X}
\put(80,235.8){$<$}\put(140,235.8){$>$}
\put(20,200){\line(1,1){40}} \put(40,220){\vector(1,1){2}}
\put(160,240){\line(1,-1){40}}\put(180,220){\vector(-1,1){2}}
\put(300,280){\line(1,-1){40}} \put(320,260){\vector(1,-1){2}}
\put(440,240){\line(1,1){40}}\put(460,260){\vector(-1,-1){2}}
\put(340,240){\line(1,0){100}}\put(390,235.6){X}
\put(360,235.8){$<$}\put(420,235.8){$>$}
\put(300,200){\line(1,1){40}} \put(320,220){\vector(1,1){2}}
\put(440,240){\line(1,-1){40}}\put(460,220){\vector(-1,1){2}}
\put(280,280){$Q_i$}\put(280,190){$Q_j$} \put(490,280) {$\bar
U_k$}\put(490,190){$\bar  D_l$} \put(0,280){$\bar D_l$}
\put(0,190){$Q_j$} \put(210,280){$\bar U_k $}\put(210,190){$Q_i$}
\put(80,250){$L_4$}\put(140,250){$\bar L_4$}
\put(360,250){$D_4$}\put(420,250){$\bar D_4 $}
\put(60,225){$y'$}\put(155,225){$y''$}\put(340,225){$\lambda^q$}\put(435,225){$\lambda''$}
\put(20,100){\line(1,-1){40}} \put(40,80){\vector(1,-1){2}}
\put(160,60){\line(1,1){40}}\put(180,80){\vector(-1,-1){2}}
\put(60,60){\line(1,0){100}}\put(110,55.6){X}
\put(80,55.8){$<$}\put(140,55.8){$>$}
\put(20,20){\line(1,1){40}}\put(40,40){\vector(1,1){2}}
\put(160,60){\line(1,-1){40}}\put(180,40){\vector(-1,1){2}}
\put(300,100){\line(1,-1){40}} \put(320,80){\vector(1,-1){2}}
\put(440,60){\line(1,1){40}}\put(460,80){\vector(-1,-1){2}}
\put(340,60){\line(1,0){100}}\put(390,55.6){X}
\put(360,55.8){$<$}\put(420,55.8){$>$}
\put(300,20){\line(1,1){40}} \put(320,40){\vector(1,1){2}}
\put(440,60){\line(1,-1){40}}\put(460,40){\vector(-1,1){2}}
\put(280,100){$\bar E_k$}\put(280,10){$\bar U_l$} \put(490,100)
{$L_i$}\put(490,10){$Q_j$} \put(0,100){$\bar E_k$}
\put(0,10){$L_i$} \put(210,100){$\bar U_l $}\put(210,10){$Q_j$}
\put(80,70){$L_4$}\put(140,70){$\bar L_4$}
\put(360,70){$D_4$}\put(420,70){$\bar D_4 $}
\put(60,45){$y$}\put(155,45){$y''$}\put(340,45){$\lambda^l$}\put(435,45){$\lambda'$}
\put(245,180){\Large $(a)$} \put(245,5){\Large  (b)}
\end{picture}
\caption{Dimension-5 operators which conserve baryon number. (b)
contains FCNC interactions. For $i\neq k$ (b) also violates lepton
number.}
\end{figure}

\begin{figure}
\begin{picture}(600,600)
\put(225,450){\line(-1,1){50}} \put(200,475){\vector(1,-1){2}}
\put(225,450){\line(1,1){50}} \put(250,475){\vector(-1,-1){2}}
\put(185,460){\Large $\bar e_k$}\put(250,460){\Large $\bar u_l$}
 \put(225,450){\line(1,-1){10}}
\put(240,435){\line(1,-1){10}} \put(255,420){\line(1,-1){10}}
\put(270,405){\line(1,-1){10}} \put(180,430){\Large $\tilde
\nu_{iL}$}\put(250,430){\Large $\tilde d'_{jL} $}
\put(215,380){\Large $\tilde w$}
\put(225,450){\line(-1,-1){10}} \put(210,435){\line(-1,-1){10}}
\put(195,420){\line(-1,-1){10}} \put(180,405){\line(-1,-1){10}}
\put(170,395){\line(1,0){110}} \put(170,395){\line(0,-1){70}}
\put(280,395){\line(0,-1){70}}
\put(170,360){\vector(0,1){2}}\put(280,360){\vector(0,1){2}}
\put(145,360){\Large $e_{i}$}\put(290,360){\Large $u_{j''}$}
\end{picture}
\vspace{-10.5cm} \caption{Dimension-5 operators dressed by wino
which mediate FCNC processes of up-type quarks.}
\end{figure}
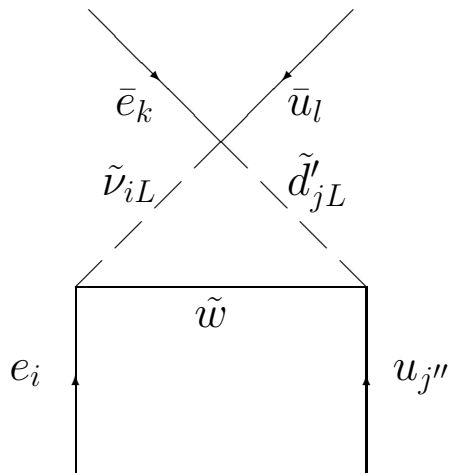

\end{document}